\documentclass{ieeeaccess}
\usepackage{cite}
\usepackage{amsmath,amssymb,amsfonts}
\usepackage{algorithmic}
\usepackage{graphicx}
\usepackage{textcomp}
\usepackage{listings}
\usepackage{tabularx}
\usepackage{multirow}
\usepackage{float}
\usepackage{hyperref}
\usepackage{subfig}
\usepackage{placeins}
\usepackage{listings}
\usepackage{enumitem}

\renewcommand{\arraystretch}{1.1} 

\newcolumntype{s}{>{\hsize=.5\hsize}X}
\newcolumntype{g}{>{\hsize=1.5\hsize}X}

\def\BibTeX{{\rm B\kern-.05em{\sc i\kern-.025em b}\kern-.08em
    T\kern-.1667em\lower.7ex\hbox{E}\kern-.125emX}}
\begin{document}
\history{Date of publication xxxx 00, 0000, date of current version xxxx 00, 0000.}
\doi{10.1109/ACCESS.2017.DOI}

\title{Large Language Models Versus Static Code Analysis Tools: A Systematic Benchmark for Vulnerability Detection}

\author{\uppercase{Damian Gnieciak}\authorrefmark{1}, 
\uppercase{Tomasz Szandala\authorrefmark{1}}
}
\address[1]{Faculty of Information and Communication Technology, Wroclaw University of Science and Technology, Wroclaw, Poland}

\markboth
{D.Gnieciak, T.Szandala \headeretal: Large Language Models Versus Static Code Analysis Tools}
{D.Gnieciak, T.Szandala \headeretal: Large Language Models Versus Static Code Analysis Tools}

\corresp{Corresponding author: Tomasz Szandala (e-mail: Tomasz.Szandala@pwr.edu.pl)}

\begin{abstract}
Modern software relies on a multitude of automated testing and quality assurance tools to prevent errors, bugs and potential vulnerabilities.
This study sets out to provide a head‑to‑head, quantitative and qualitative evaluation of six automated approaches: three industry-standard rule‑based static code‑analysis tools (SonarQube, CodeQL and Snyk~Code) and three state‑of‑the‑art large language models hosted on the GitHub Models platform (GPT‑4.1, Mistral~Large and DeepSeek~V3).
 Using a curated suite of ten real‑world C\# projects that embed 63 vulnerabilities across common categories such as SQL injection, hard‑coded secrets and outdated dependencies, we measure classical detection accuracy (precision, recall, F‑score), analysis latency, and the developer effort required to vet true positives.
The language‑based scanners achieve higher mean F‑1 scores,0.797, 0.753 and 0.750, than their static counterparts, which score 0.260, 0.386 and 0.546, respectively. LLMs' advantage originates from superior recall, confirming an ability to reason across broader code contexts. However, this benefit comes with substantial trade‑offs: DeepSeek~V3 exhibits the highest false‑positive ratio, all language models mislocate issues at line‑or‑column granularity due to tokenisation artefacts.
Overall, language models successfully rival traditional static analysers in finding real vulnerabilities. Still, their noisier output and imprecise localisation limit their standalone use in safety‑critical audits. We therefore recommend a hybrid pipeline: employ language models early in development for broad, context‑aware triage, while reserving deterministic rule‑based scanners for high‑assurance verification. The open benchmark and JSON‑based result harness released with this paper lay a foundation for reproducible, practitioner‑centric research into next‑generation automated code security.

ProjectAnalyzer Code is available on GitHub:\url{https://github.com/Damian0401/ProjectAnalyzer}
\end{abstract}


\titlepgskip=-15pt

\maketitle

\section{Introduction}\label{sec:introduction} 
Rapidly developed software has become vital for modern industry. Yet, its growing complexity and attack surface have outpaced the human capacity to reason about every line of code~\cite{ajiga2024role}. Static Application Security Testing (SAST) has long been the developer's first line of defence: tools such as SonarQube, CodeQL, and SnykCode integrate seamlessly into continuous‑integration pipelines, flagging defects before they ever reach production, and are trusted by more than seven~million developers across 400,000 organisations worldwide. 

Recent empirical studies nevertheless reveal a persistent gap, while a single state‑of‑the‑art analyser can highlight vulnerabilities in roughly half of real‑world, vulnerability‑contributing commits, false‑positive noise and rule coverage limitations continue to hamper day‑to‑day adoption.
Concurrently, the remarkable rise of large language models (LLMs) has opened a new branch of code intelligence. Industry surveys chart a rapid uptick in LLM pilots across engineering teams. The research community has begun to fine‑tune these models for vulnerability detection with promising results: early experiments report competitive F‑scores against graph‑based and traditional sequence models, while comparative analyses on emerging models—GPT‑4‑class systems, Mistral‑Large, DeepSeek~V3 and others—suggest that LLMs can reason across larger contexts and identify subtle data‑flow flaws that evade conventional pattern‑based rules.

Yet, enthusiasm must be tempered with caution. The same generative power that allows an LLM to "understand" source code also enables failure modes alien to deterministic analysers. For instance, a recent industry report warns of "slopsquatting": hallucinated package names inserted by LLMs that attackers later register in public repositories, creating an entirely new supply‑chain vector. Such findings underline the need for a principled, head‑to‑head assessment rather than anecdotes and marketing claims.TechRadar

This paper undertakes the first systematic comparison between three mature static analysers, SonarQube, CodeQL and SnykCode, and three cutting‑edge LLMs, available on GitHub: GPT‑4.1, Mistral‑Large and DeepSeek~V3, evaluated on a standard benchmark suite spanning multiple projects and vulnerability classes. We measure classical accuracy metrics (precision, recall, F1), as well as developer‑centric costs, such as inspection effort per true positive, time‑to‑signal in CI, and the qualitative value of the explanations generated by each approach.

Our contributions consits of:
\begin{itemize}
    \item Unified benchmark and protocol. We form a diverse dataset of real‑world vulnerable functions, pair each with ground‑truth labels, and release an open evaluation harness suitable for rule‑based and generative systems.
    \item Comprehensive quantitative evaluation. We show where LLMs already rival, and occasionally surpass, specialist analysers in recall, while highlighting scenarios in which deterministic rules remain indispensable for suppressing false alarms.
    \item Qualitative insight. Through thematic analysis of model rationales and tool reports, we surface strengths (e.g., data‑flow reasoning across files) and weaknesses (e.g., inconsistent CWE mapping) unique to each paradigm.
    \item Guidance for practitioners and researchers. We distil our findings into recommendations on tool choice, prompt engineering, and future benchmarks that capture emerging threat vectors.
\end{itemize}
\section{State of the Art}

Static code analysis represents an essential component within modern software development processes. Early detection of software defects, which, according to various studies, can lead to a significant reduction in the cost and impact of subsequent corrections \cite{Kaestner2018, Venkatasubramanyam2015}. Furthermore, it simplifies the code review process by assisting developers in detecting potential issues early, reducing the time and effort required for manual code reviews \cite{reduceCost2017Singh}. It is worth paying attention to the fact that the utility of static analysis is not limited to professional and commercial use. Academic research highlights its educational value. A~large-scale study analysing more than 500 student-developed software projects illustrates that regular static code analysis tools can significantly improve code quality and student learning outcomes \cite{Nikolic2024}. With the increasing integration of artificial intelligence into the software development lifecycle, the role and application of static code analysis are also poised to evolve. AI-powered development tools are a way of reinventing traditional programming practices and introducing new possibilities. This approach will enhance the effectiveness of static analysis techniques and redefine their function, enabling more contextualised, personalised, and proactive code evaluation processes. 

\subsection{How Static Code Analysis Works}

Static source code analysis involves examining the source code without executing it to detect potential bugs and weaknesses. The code is analysed for defined patterns or rules that indicate undesirable behaviour. Typical tools use control, data flow analysis, and pattern matching to known error signatures \cite{bardas2010static}. Static scanners can detect, e.g. SQL Injection or code injection vulnerabilities through taint analysis, but usually only if such patterns have been predefined in their rules. In industry practice, rule sets are constantly expanded to include new bugs. Due to the multiplicity of use cases, they have been divided into categories \cite{taxonomy2010Novak}, such as:

\begin{itemize}
\item \textbf{Naming} – Issues related to inconsistent, misleading, or nonstandard naming conventions for identifiers.
\item \textbf{Style} – Violations of code formatting or stylistic guidelines can affect readability.
\item \textbf{Concurrency} – Problems arising from incorrect handling of parallel execution or shared resources.
\item \textbf{Exceptions} – Improper use or handling of exceptions and error conditions.
\item \textbf{Performance} – Inefficiencies or bottlenecks that could degrade the system's performance.
\item \textbf{Interoperability} – Compatibility problems between different systems, platforms, or APIs.
\item \textbf{Security} – Vulnerabilities that could be exploited to compromise the system's integrity or data.
\item \textbf{Maintainability} – Code complexity or structure issues that make future modification or understanding more difficult.
\item \textbf{General} – Broad or uncategorised issues that do not fit neatly into other specific groups.
\end{itemize}

\subsection{Format of Results}

Static code analysis tools are often integrated into other software development processes. The analysis results are usually an essential element before the final implementation of the system under development.
\begin{figure*}[!t]                 
  \centering                        
\begin{lstlisting}[caption={Example SARIF JSON output},captionpos=b,label={listing:sarif_json}]
{
"$schema": "https://json.schemastore.org/sarif-2.1.0.json",
"version": "2.1.0",
"runs": [{
"tool": {
  "driver": {
    "name": "SonarQube",
    "semanticVersion": "1.0.0",
    "version": "1.0.0",
    "rules": [{
      "id": "SQL_INJECTION",
      "shortDescription": { "text": "SQL Injection vulnerability" }
    }]
  }
},
"results": [{
  "ruleId": "SQL_INJECTION",
  "ruleIndex": 0,
  "level": "error",
  "message": { "text": "User input is used directly in a SQL query. This can lead to SQL injection." },
  "locations": [{
    "physicalLocation": {
      "artifactLocation": { "uri": "Controllers/UserController.cs" },
      "region": {
        "startLine": 27,
        "endLine": 27,
        "startColumn": 17,
        "endColumn": 66
      }
    }
  }]
}]
}]
}
\end{lstlisting}
\end{figure*}

SARIF (Static Analysis Results Interchange Format) was presented in 2020 \cite{sarif210} as a~standardized and extensible format for exchanging results produced by static analysis tools, enabling consistent interpretation, comparison, and integration between different platforms. Since then, it has been adopted and integrated by many static analysis tools \cite{kummita2019sarif}. The structure shown in  listing~\ref{listing:sarif_json} represents a simplified example of a SARIF JSON output. The list includes only the fields necessary to generate a minimal report.

\begin{itemize}[leftmargin=*]

\item[] \textbf{\$schema}: Specifies the URL of the JSON schema used to validate the structure, referencing the official SARIF 2.1.0 specification.
\item[] \textbf{version}: Defines the version of the SARIF format that is being used (here, \textit{"2.1.0"}).
\item[] \textbf{runs}: An array containing analysis runs.
\item[] \textbf{tool}: Provides metadata on the analysis tool.
\item[] \textbf{driver}: Describes the tool driver (e.g., SonarQube).
\item[] \textbf{name}, \textbf{semanticVersion}, \textbf{version}: Identify the tool and its version.
\item[] \textbf{rules}: An array of rules-objects applied during analysis.
\item[] \textbf{id}: Unique identifier for the rule (e.g., \textit{"SQL\_INJECTION"}).
\item[] \textbf{shortDescription}: A brief text describing the rule.

\item[] \textbf{results}: An array of analysis results.
\item[] \textbf{ruleId}, \textbf{ruleIndex}: Reference the rule responsible for the finding.
\item[] \textbf{level}: Indicates the severity level of the result (e.g., \textit{"error"}).
\item[] \textbf{message}: A descriptive message on the issue was found.
\item[] \textbf{locations}: An array indicating the location of the issue in the code.
\item[] \textbf{physicalLocation}: Provides detailed location data.

\item[] \textbf{artifactLocation}: The file where the issue occurs.
\item[] \textbf{region}: Describes the exact range in the file (line and column numbers).

\end{itemize}

This structure is not a complete representation of all available SARIF fields. According to the official SARIF schema~\cite{sarif-schema}, many additional fields and features are supported. The example includes only the fields required to produce a minimal report.

\subsection{Limitations of Traditional Code Analysis}

Despite their widespread use, static approaches face significant challenges. First, designing a good analyser is time-consuming and difficult; each new type of defect requires the development of an appropriate rule. As a result, typical tools are limited to predefined defect patterns and may fail to recognise new or unusual problems \cite{liang2012codas}. Another limitation may be the lack of ability to fully understand the context of the analysed code, as shown by the authors of the paper \cite{simoes2024evaluating}. This shortcoming underscores the need for more adaptive and intelligent approaches and methods to fill the gap between recognising a~known pattern and deep semantic understanding.

\subsection{Use of Large Language Model}

In recent years, numerous works have examined large language models in the context of  industry code quality \cite{ahmad2025lashed,szandala2025chatgpt}. Models trained on huge collections of code and text demonstrate the ability to understand the context of a program and suggest corrections. This raises the question of whether they can take over the role of classic static analysers in bug detection. An important advantage of LLMs is the lack of rigidly defined rules - the model can potentially catch an error based on its "understanding" of the context, even if the pattern is not explicitly programmed. Research shows that LLMs can detect certain defects without the need for a complete set of tests or rules, due to their ability to infer from the context of the code. The comparative potential of large language models and traditional static analysis tools has been explored in studies such as \cite{simoes2024evaluating}, which analysed the performance of GPT-3.5 Turbo and GPT-4o in relation to SonarQube. This work is extended by additional static analysers, including CodeQL and SnykCode, as well as a broader set of large language models such as GPT-4.1, DeepSeek~V3, and Mistral Large, enabling a more comprehensive evaluation of their capabilities in the detection of defects in the source code.
\section{Methods}

Ensuring code quality and reliability is crucial in modern IT systems. Static analysis tools typically identify vulnerabilities based on predefined patterns or rule sets. The effectiveness of such tools in improving software quality and their efficiency in minimising code defects has been demonstrated in the existing literature \cite{nichols_static_analysis_2020}. 

\subsection{Overview of Evaluated Tools} 

For this work, three widely recognised static analysis tools have been selected for evaluation: SonarQube, SnykCode and CodeQL.

SonarCode is an open source tool that is used for static code analysis, created in 2006 \cite{wikipediaSonarQube}. It is designed to automatically detect errors, security vulnerabilities, and issues related to code quality and technology debt. During almost 20 years of existence on the market, it has grown to support over 30 programming languages \cite{sonarqubeDocsLanguages} and offers a built-in interface with comprehensive dashboards and detailed reports. This tool is used by more than 400,000 organisations \cite{sonarsource2024}. When choosing SonarQube, it is possible to select one of two options to use this solution. 

The user can self-host the Community version, which is open source and can be used initially without any costs. In that case, it is in the consumer's interests to ensure the safety of processed data. The second option is to choose one of the two paid versions provided by SonarSource, which are Team and Enterprise \cite{sonarqubePricing}. The first version costs 65\$ per month (as of 2025), and the cost of the higher tier is determined individually after contacting the sales department. SonarSource guarantees the respect of the EU General Data Protection Regulation (GDPR) or the California Consumer Privacy Act to ensure the privacy of their users' data \cite{sonarqubeDataPrivacy}. Due to its long existence on the market, SonarCode is a~very mature and proven tool.

CodeQL is an engine created and developed by GitHub. Its main goal is discovering security vulnerabilities and bugs in the analysed source code. The distinguishing feature of this tool over others is that it allows users to write custom queries using a specialised domain-specific language named QL. CodeQL is part of the GitHub Advanced Security platform and can be used with most popular programming languages, such as C++, Java, and Python \cite{githubCodeQLDocs}. CodeQL, as part of GitHub Code Security, also offers two pricing levels, Team for the price of 4\$ per month for each user and Enterprise starting at 21\$ per month for each user \cite{codeQlPricing}. GitHub processed personal data according to the declaration of respect for the GDPR and other applicable laws \cite{codeQlDataPrivacy}.

SnykCode is a static application security testing tool created by the Snyk company. Unlike traditional SAST tools, SnykCode uses machine learning algorithms to detect issues in the source code \cite{snykCodeDocs}. This tool focuses mainly on security vulnerabilities such as SQL injection, cross-site scripting (XSS), and authentication-related issues. The first version was officially released in 2020 \cite{snykCodeRelease}. SnykCode is also available in two pricing plans, Team for 25\$ per month for each contributing developer and Enterprise, where the price is determined individually after contacting the sales department, similar to SonarQube \cite{snykCodePricing}. To ensure data privacy, Snyk leads a global program designed to align with the requirements of the GDPR and other privacy laws \cite{snykCodeDataPrivacy}.

Before using each static analysis tool, it is crucial to be familiar with the license and data privacy arrangements. Another important aspect is to conduct a cost analysis. Key aspects were visualised with a Table~\ref{table:static_analysis_tools_comprasion} to compare each tool better. All three solutions comply with data privacy regulations and offer multiple pricing tiers, but differ in hosting models and cost structures, which may influence the selection based on organisational needs.

\begin{table*}[t]
\centering
\renewcommand{\arraystretch}{1.3}
\begin{tabularx}{\textwidth}{sXXX}
& \textbf{SonarQube} & \textbf{CodeQL} & \textbf{SnykCode} \\
\hline
Hosting model & Self-hosted (Community) or cloud-based (Team, Enterprise) & Cloud-based via GitHub Code Security & Cloud-based (Team, Enterprise) \\
\hline
Cost of license & Free (Community), \$65/month (Team), Enterprise (custom pricing) & \$4/user/month (Team), from \$21/user/month (Enterprise) & \$25/user/month (Team), Enterprise (custom pricing) \\
\hline
Data privacy & Compliant with GDPR and CCPA & Compliant with GDPR and other applicable laws & Global program aligned with GDPR and other privacy laws \\
\hline
\end{tabularx}
\caption{Comparison of Code Analysis Tools}
\label{table:static_analysis_tools_comprasion}
\end{table*}

\subsection{Large Language Models in Code Analysis}

The use of large language models can significantly improve code quality assessments, as they can analyse predefined rules and vulnerabilities and contextual aspects of the code under review \cite{simoes2024evaluating}. For this work, three large language models from leading companies have been
selected for evaluation: GPT 4.1, DeepSeek V3, and Mistral Large.

GPT 4.1, developed by OpenAI, is known for its extensive context window of 1,047,576 tokens, significantly enhancing its ability to handle complex tasks, such as analysing large codebases or extensive documents. The price is set at \$2 per million input tokens and \$8 per million output tokens \cite{gptPricing}. Regarding data privacy, user interactions with GPT 4.1 in consumer products can be used to train OpenAI models, but users can opt out through settings. For business products, data is not used for training by default \cite{gptDataPrivacy}. The declared knowledge cut-off date is June 2024 \cite{gptKnowledgeCutoff}.

DeepSeek V3 offers a cost-effective alternative with a context window of 64,000 tokens, attractively priced at \$0.27 per million input tokens and \$1.10 per million output tokens \cite{deepSeekPricing}. However, DeepSeek collects extensive personal data, including user inputs, uploaded files, and automatically collected network information such as IP addresses and device identifiers, raising significant concerns about data privacy \cite{deepSeekDataPrivacy}. The declared knowledge cut-off date is July 2024 \cite{deepSeekKnowledgeCutoff}. It was created by the Chinese company DeepSeek, founded in 2023. The model has gained popularity due to its competitive capabilities in relation to the cost of its exploitation.

Mistral Large, from Mistral AI, features a balanced context window of 32,000 tokens and costs \$2 per million input tokens and \$6 per million output tokens \cite{mistralPricing}. Its data privacy policy states that user data is not utilised for training except in specific cases such as free-tier usage without explicit opt-out, feedback provision, or content moderation purposes \cite{mistralDataPrivacy}.

Before integrating LLM into practical workflows, it is essential to understand their licensing terms, data privacy policies, and associated costs. These factors are critical in selecting a model that aligns with organisational requirements. Key aspects were visualised with a Table~\ref{table:large_language_models_comprasion} to compare each LLM better. Each of the selected models has distinct advantages and limitations. GPT 4.1 excels with its unparalleled context capacity, which is essential for complex tasks, such as analysing a large codebase, but it has higher pricing. DeepSeek V3 offers significant cost benefits, but at the expense of increased data privacy risks. Mistral Large presents a similar pricing and approach to data privacy to GPT 4.1, but is slightly more cost-effective.

\begin{table*}
\centering
\renewcommand{\arraystretch}{1.3} 
\begin{tabularx}{\textwidth}{sXXX}
 & \textbf{GPT 4.1} & \textbf{DeepSeek V3} & \textbf{Mistral Large} \\
\hline
Knowledge cutoff & June 1, 2024 & July 1, 2024 & Unknown \\
\hline
Context window (in tokens) & 1,047,576 & 64,000 & 32,000 \\
\hline
Cost of usage (per 1M tokens) & 2\$ input, 8\$ output & 0.27\$ input, 1.1\$ output & 2\$ input, 6\$ output \\
\hline
Data privacy &  User data may be used for training (opt-out available for consumers; not used by default for business products). & Extensive data collection including inputs, chat history, and automated network data collection. & Data used for training only under specific conditions (e.g., free tier without opt-out, moderation purposes). \\
\hline
\end{tabularx}
\caption{Comparison of Large Language Models}
\label{table:large_language_models_comprasion}
\end{table*}

The effective use of large language models for static analysis requires a structured approach to data preparation and result handling. This section covers the process of selecting and preparing software projects, describes the SARIF format used to represent results, and presents the tool created to perform the analysis.

\subsection{Dataset}

For the purpose of analysis, 10 projects were prepared in the popular programming language, which is C\#. According to Table~\ref{table:project_statistics}, size of the prepared projects is between 3500 and 5500 characters and contains up to 10 files. This number of files was selected based on the analysis, which shows that it is the most common change based on the source control history~\cite{ferreira2022commits}.

\begin{table}
\centering
\begin{tabularx}{\linewidth}{cccc}
\hline
\textbf{Project} & \textbf{Number of files} & \textbf{Number of characters} & \textbf{Number of vuln.} \\
\hline
S01 & 5 & 5317 & 8 \\
S02 & 3 & 3145 & 3 \\
S03 & 3 & 3431 & 6 \\
S04 & 4 & 4016 & 1 \\
S05 & 3 & 5117 & 6 \\
S06 & 7 & 3791 & 8 \\
S07 & 7 & 5036 & 13 \\
S08 & 8 & 3878 & 7 \\
S09 & 7 & 3709 & 8 \\
S10 & 10 & 4938 & 3 \\
\hline
\end{tabularx}
\caption{Number of files, characters and vulnerabilities in each project}
\label{table:project_statistics}
\end{table}

Each of project has a random number of vulnerabilities, such as:

\begin{itemize}
    \item SQL injection,
    \item cross-site scripting,
    \item hardcoded secrets,
    \item command injection,
    \item weak encryption algorithms,
    \item deprecated dependencies.
\end{itemize}

\subsection{ProjectAnalyzer}
Working with large language models requires intensive computational power, which presents a notable challenge for consistent evaluation. Furthermore, GPT 4.1, one of the three models originally intended for study, is not publicly accessible. As a result, to ensure equivalent conditions and a fair basis for comparison, the GitHub Models platform was chosen as the hosting environment. GitHub Models is a publicly accessible repository developed through a collaboration between GitHub and Microsoft \cite{github-models}, providing a standardized and open place to prototype and build AI-powered solutions.

For the purpose of generating reports in SARIF format, the tool named \textit{ProjectAnalyzer} was developed. It is designed to automate the analysis of software projects using large language models. The tool facilitates static analysis by scanning the current working directory for source code files with specified extensions (e.g., \textit{cs}, \textit{.csproj}, \textit{.sln}). Once the relevant files are identified, the tool constructs a single aggregated prompt that represents the entire project, suitable for processing. The prompt is then sent by API to each of the selected models. Each model processes the prompt independently and returns an analysis in JSON format. This response is subsequently parsed and transformed into the SARIF format. The final report is saved to the specified output location, providing a machine-readable summary of the models findings. C\# was chosen as the programming language to develop the analyzer due to its native and officially supported capabilities for interacting with the API used in this project \cite{microsoft_extensions_ai_2025}. 

\begin{figure}[H]
    \centering
    \includegraphics[width=.9\linewidth]{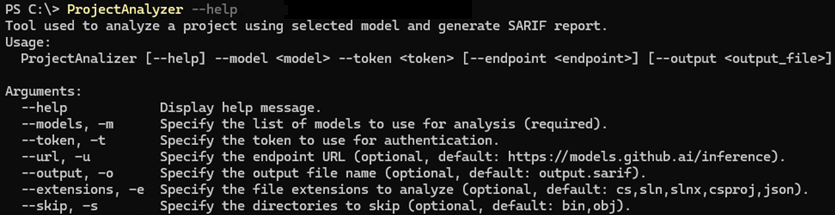}
    \caption{ProjectAnalyzer - help command}
    \label{fig:tool-help}
\end{figure}

\begin{figure}[H]
    \centering
    \includegraphics[width=.9\linewidth]{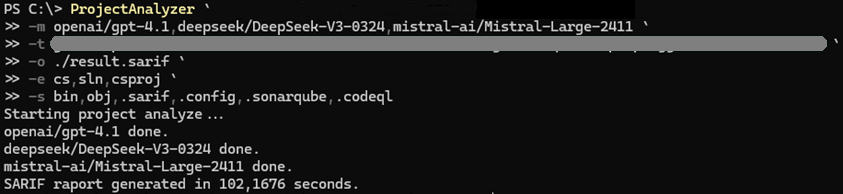}
    \caption{ProjectAnalyzer - usage}
    \label{fig:tool-usage}
\end{figure}

During analysis, a system prompt, shown in Figure~\ref{listing:prompt}, was included as a system message. It was intended to instruct the model on how it should behave and in what format the results should be returned. Figure~\ref{fig:tool-help} illustrates the help command of the \textit{ProjectAnalyzer} tool, displaying available arguments, flags, and usage options, while Figure~\ref{fig:tool-usage} illustrates the usage of the tool, showing the summary of the completed analyzes for each selected model, and reporting the time required to generate the full report. The token used to access models from the GitHub Models platform was hidden. 
\begin{figure*}[!t]
  \centering 
\begin{lstlisting}[caption={Prompt sent used for analysis},captionpos=b,label={listing:prompt}]

You are a static code analysis engine. Your task is to review the provided source code and identify security vulnerabilities. Focus on detecting vulnerabilities such as (but not limited to):
- SQL Injection
- Cross-Site Scripting (XSS)
- Command Injection
- Insecure Deserialization
- Insecure or missing authentication/authorization mechanisms
- Hardcoded credentials or secrets
- Improper input validation or lack of sanitization
- Use of outdated or vulnerable libraries
- Insecure use of cryptography (e.g., weak algorithms, hardcoded keys)
- Insecure file handling (e.g., path traversal, unrestricted uploads)

Only analyze and report issues that pose a security risk. Do not report code smells, general bugs, or non-security-related issues.
Your output must be a JSON array, enclosed between triple backticks (```json and ```), with each finding represented as a JSON object in the following format:
[{"RuleId":"string","RuleDescription":"string","Level":"Error"|"Warning"|"Note"|"None", "Message":"string","Path":"string","Category":"string","StartLine":integer, "EndLine":integer,"StartColumn":integer,"EndColumn":integer}]

Field description:
- RuleId: A short identifier for the rule or issue.
- RuleDescription: A brief description of the rule being violated.
- Level: Severity of the issue (Error, Warning, Note, or None).
- Message: A concise explanation of the specific issue found.
- Path: The file path where the issue occurs.
- Category: The general category.
- StartLine, EndLine: Line range of the issue.
- StartColumn, EndColumn: Column range of the issue.

Ensure the JSON is well-formed and strictly adheres to this json structure. All fields are required.

\end{lstlisting}
\end{figure*}

\section{Results}

A set of standardised ML model performance metrics was utilised for each static analysis tool and the large language model under consideration to evaluate all the projects prepared. Specifically, the evaluation measured the execution time, the number of detected vulnerabilities, and the number of true positive identifications. These indicators form the basis for a more in-depth and comparative analysis.

\subsection{Definition of Evaluation Metrics}

The total number of identified vulnerabilities was determined by counting the number of entries present in the Results section of the SARIF reports generated by the analysis tools \cite{sarif-schema}. Each entry in this section represents a distinct issue identified during the analysis process, and the aggregate count serves as a quantitative indicator of detection capabilities.

Each reported vulnerability was individually reviewed and compared with a reference dataset of known vulnerabilities for the given project to determine the number of true positive results. Only findings that accurately matched a real problem were counted as true positives. The only exception to this rule was the location region of the results produced by the large language models. The rationale behind this exception is explained in detail in the Discussion section.

The number of false positive results was determined by subtracting the number of confirmed true positives from the total number of vulnerabilities identified by each tool or model. A false positive is defined as a reported vulnerability that does not correspond to any real or known problem in the reference data set for the analysed project.

The F1 score is a standard evaluation metric used to assess the overall effectiveness of a classification model by combining precision and recall into a single harmonic mean. In the context of vulnerability detection, it provides a balanced view of how well a tool or model identifies true vulnerabilities while minimising incorrect detections. To fully understand and calculate the F1 score, it is essential to introduce the individual metrics on which it depends, including false negatives, precision, and recall, which together determine the overall value of the score.

\textbf{False negatives} is calculated by subtracting the number of true positive results from the total number of known vulnerabilities in the reference data set for each project.

\[
\text{False Negatives} = \text{Total Known Vulnerabilities} - \text{True Positives}.
\]

\textbf{Precision} is defined as the proportion of correctly identified vulnerabilities (True Positives) among all reported vulnerabilities and is calculated as:

\[
\text{Precision} = \frac{\text{True Positives}}{\text{True Positives} + \text{False Positives}}.
\]

\textbf{Recall}, on the other hand, measures the proportion of actual vulnerabilities that were correctly identified by the tool and is given by:

\[
\text{Recall} = \frac{\text{True Positives}}{\text{True Positives} + \text{False Negatives}}.
\]

The \textbf{F1 score} is then calculated as the harmonic mean of precision and recall, providing a single measure that balances both:

\[
\text{F1 Score} = 2 \cdot \frac{\text{Precision} \cdot \text{Recall}}{\text{Precision} + \text{Recall}}.
\]

This metric is handy when comparing tools with varying balances of precision and recall, as it offers a comprehensive view of detection performance. All F1 scores reported in this work were calculated based on manually verified true positive results.

Finally, the execution time was systematically measured for each run of the static analysis tools and the implemented analyser. These measurements were recorded in seconds, maintaining a precision of three floating-point decimal places, to support accurate performance comparisons and detailed evaluation.

\subsection{Experiments}
\label{sec:experiments}

The results presented in the following tables illustrate the raw performance data gathered during these experiments. This data is then further analysed to provide insight into each method's strengths, limitations, and overall reliability. The experiments were run in a uniform execution environment to eliminate variability due to system performance. Furthermore, each static analysis tool and large language model was executed using their default configuration settings, unless explicitly stated otherwise, to maintain fairness and reproducibility in all approaches tested. 

Table~\ref{table:tool_statistics} presents the total number of vulnerabilities found and the number of true positive results for the static analysis tools. In contrast, Table~\ref{table:model_statistics} provides the corresponding data for the large language models. 
It should be noted that \textit{SnykCode} consistently reports a significantly higher number of vulnerabilities compared to the other tools evaluated. This trend may indicate a higher detection sensitivity, suggesting that \textit{SnykCode} employs a more aggressive strategy to identify potential problems.

Table~\ref{table:execution_time} shows the total execution time measured for both the large language models and the static analysis tools. These measurements were obtained following the procedure described in previous section. It should be noted that the execution time recorded for \textit{CodeQL} is significantly longer than that of the other tools, which may reflect its deeper or more comprehensive analysis process.
\begin{table}[]
\centering
\begin{tabularx}{\linewidth}{X|XX|XX|XX}
\hline
Project & SonaQube Total & SonaQube TP & CodeQL Total & CodeQL TP & Snyk Total & Snyk TP \\
\hline
S01 & 6 & 4 & 6 & 3 & 10 & 7 \\
S02 & 1 & 1 & 1 & 1 & 1  & 1 \\
S03 & 2 & 2 & 3 & 3 & 3  & 3 \\
S04 & 0 & 0 & 0 & 0 & 0  & 0 \\
S05 & 0 & 0 & 3 & 3 & 1  & 1 \\
S06 & 1 & 1 & 3 & 3 & 7  & 7 \\
S07 & 1 & 1 & 5 & 5 & 14 & 10 \\
S08 & 0 & 0 & 1 & 0 & 7  & 5 \\
S09 & 4 & 4 & 4 & 3 & 11 & 8 \\
S10 & 0 & 0 & 0 & 0 & 3  & 0 \\
\hline
\end{tabularx}
\caption{Number of Total Found and True Positive~(TP) findings for each project and tool}
\label{table:tool_statistics}
\end{table}

\begin{table}[]
\centering
\begin{tabularx}{\linewidth}{X|XX|XX|XX}
\hline
Project & GPT-4.1 Total & GPT-4.1 TP & Mistral Total & Mistral TP & DSeek Total & DSeek TP \\
\hline
S01 & 13 & 8 & 11 & 7 & 11 & 7 \\
S02 &  3 & 3 &  3 & 2 &  3 & 3 \\
S03 &  8 & 6 &  8 & 6 &  8 & 6 \\
S04 &  2 & 1 &  3 & 1 &  5 & 1 \\
S05 &  4 & 4 &  5 & 4 &  6 & 5 \\
S06 & 10 & 7 &  8 & 6 &  8 & 6 \\
S07 & 15 &11 &  9 & 8 & 10 & 8 \\
S08 &  7 & 5 &  5 & 5 &  7 & 6 \\
S09 &  9 & 8 &  7 & 7 &  7 & 7 \\
S10 &  3 & 2 &  2 & 2 &  5 & 3 \\
\hline
\end{tabularx}
\caption{Number of Total Found and True Positive~(TP) per project and model}
\label{table:model_statistics}
\end{table}

\begin{table*}[h!]
\centering
\begin{tabularx}{\textwidth}{XXXXXXXX}
\multirow{2}{*}{\textbf{Project}} & 
\multirow{2}{*}{\textbf{SonarQube}} & 
\multirow{2}{*}{\textbf{CodeQL}} & 
\multirow{2}{*}{\textbf{SnykCode}} & 
\multirow{2}{*}{\textbf{GPT-4.1}} & 
\multirow{2}{*}{\textbf{Mistral}} & 
\multirow{2}{*}{\textbf{DeepSeekV3}} \\
& & & & & & \\
\hline
S01 & 39.928s & 195.102s & 27.370s & 12.741s & 35.149s & 17.835s \\
S02 & 32.984s & 186.703s & 20.887s & 4.407s & 11.178s & 5.815s \\
S03 & 33.544s & 188.552s & 27.071s & 8.140s & 24.019s & 11.607s \\
S04 & 35.511s & 207.137s & 13.921s & 3.526s & 10.181s & 7.345s \\
S05 & 34.426s & 207.640s & 16.886s & 6.715s & 14.790s & 7.472s \\
S06 & 30.202s & 197.323s & 18.113s & 9.536s & 22.919s & 29.860s \\
S07 & 35.921s & 199.423s & 21.590s & 17.802s & 26.623s & 13.581s \\
S08 & 32.335s & 185.923s & 16.547s & 7.633s & 15.669s & 12.701s \\
S09 & 38.701s & 191.757s & 19.025s & 8.196s & 22.696s & 9.931s \\
S10 & 63.905s & 213.088s & 29.266s & 4.769s & 67.273s & 9.083s \\
\hline
\end{tabularx}
\caption{Execution time for each tool and model}
\label{table:execution_time}
\end{table*}

\section{Analysis of Results}
\label{sec:analysis_results}

As part of the effort to perform a comprehensive comparative analysis using key evaluation metrics, including False Positives, False Negatives, Precision, Recall, and the F1 Score. These metrics were computed for all large language models and static analysis tools to facilitate an objective and balanced comparison of their detection capabilities and overall effectiveness.

The detailed results of this evaluation are presented in Tables~\ref{table:sonarqube_stats}-\ref{table:deepseek_v3_stats}. The detailed results of this evaluation are presented in Tables~\ref{table:sonarqube_stats}–\ref{table:deepseek_v3_stats}. Tool-based methods include \textit{SonarQube} (Table~\ref{table:sonarqube_stats}), \textit{CodeQL} (Table~\ref{table:codeql_stats}), and \textit{SnykCode} (Table~\ref{table:snykcode_stats}), while model-based approaches include \textit{GPT-4.1} (Table~\ref{table:gpt4_1_stats}), \textit{Mistral Large} (Table~\ref{table:mistral_large_stats}), and \textit{DeepSeek V3} (Table~\ref{table:deepseek_v3_stats}). These tables compile the computed metrics for each tool across all target projects, enabling a direct side-by-side comparison of their effectiveness. The tabulated data form the basis for the comparative analysis discussed in the following sections.

The static analysis tools - \textit{SonarQube}, \textit{CodeQL}, and \textit{SnykCode} - achieved average F1 scores of 0.260, 0.386, and 0.546, respectively. In comparison, the large language models evaluated, \textit{GPT-4.1}, \textit{Mistral Large} and \textit{DeepSeek V3}, obtained average F1 scores of 0.797, 0.753, and 0.750, respectively. Detailed average results are presented in Table~\ref{table:avg_metrics_summary}.

\begin{table}[]
\centering
\begin{tabularx}{\linewidth}{XXXXXX}
\multicolumn{6}{c}{\textbf{SonarQube}} \\
\hline
\textbf{Project} & \textbf{False Positive} & \textbf{False Negative} & \textbf{Precision} & \textbf{Recall} & \textbf{F1 Score} \\
\hline
S01 & 2 & 4 & 0.667 & 0.500 & 0.571 \\
S02 & 0 & 2 & 1.000 & 0.333 & 0.500 \\
S03 & 0 & 4 & 1.000 & 0.333 & 0.500 \\
S04 & 0 & 1 & 0.000 & 0.000 & 0.000 \\
S05 & 0 & 6 & 0.000 & 0.000 & 0.000 \\
S06 & 0 & 7 & 1.000 & 0.125 & 0.222 \\
S07 & 0 & 12 & 1.000 & 0.077 & 0.143 \\
S08 & 0 & 7 & 0.000 & 0.000 & 0.000 \\
S09 & 0 & 4 & 1.000 & 0.500 & 0.667 \\
S10 & 0 & 3 & 0.000 & 0.000 & 0.000 \\
\hline
\end{tabularx}
\caption{Calculated metrics using SonarQube}
\label{table:sonarqube_stats}
\end{table}

\begin{table}[]
\centering
\begin{tabularx}{\linewidth}{XXXXXX}
\multicolumn{6}{c}{\textbf{CodeQL}} \\
\hline
\textbf{Project} & \textbf{False Positive} & \textbf{False Negative} & \textbf{Precision} & \textbf{Recall} & \textbf{F1 Score} \\
\hline
S01 & 3 & 5 & 0.500 & 0.375 & 0.429 \\
S02 & 0 & 2 & 1.000 & 0.333 & 0.500 \\
S03 & 0 & 3 & 1.000 & 0.500 & 0.667 \\
S04 & 0 & 1 & 0.000 & 0.000 & 0.000 \\
S05 & 0 & 3 & 1.000 & 0.500 & 0.667 \\
S06 & 0 & 5 & 1.000 & 0.375 & 0.545 \\
S07 & 0 & 8 & 1.000 & 0.385 & 0.556 \\
S08 & 1 & 7 & 0.000 & 0.000 & 0.000 \\
S09 & 1 & 5 & 0.750 & 0.375 & 0.500 \\
S10 & 0 & 3 & 0.000 & 0.000 & 0.000 \\
\hline
\end{tabularx}
\caption{Calculated metrics using CodeQL}
\label{table:codeql_stats}
\end{table}

\begin{table}[]
\centering
\begin{tabularx}{\linewidth}{XXXXXX}

\multicolumn{6}{c}{\textbf{SnykCode}} \\
\hline
\textbf{Project} & \textbf{False Positive} & \textbf{False Negative} & \textbf{Precision} & \textbf{Recall} & \textbf{F1 Score} \\
\hline
S01 & 3 & 1 & 0.700 & 0.875 & 0.778 \\
S02 & 0 & 2 & 1.000 & 0.333 & 0.500 \\
S03 & 0 & 3 & 1.000 & 0.500 & 0.667 \\
S04 & 0 & 1 & 0.000 & 0.000 & 0.000 \\
S05 & 0 & 5 & 1.000 & 0.167 & 0.286 \\
S06 & 0 & 1 & 1.000 & 0.875 & 0.933 \\
S07 & 4 & 3 & 0.714 & 0.769 & 0.741 \\
S08 & 2 & 2 & 0.714 & 0.714 & 0.714 \\
S09 & 3 & 0 & 0.727 & 1.000 & 0.842 \\
S10 & 3 & 3 & 0.000 & 0.000 & 0.000 \\
\hline
\end{tabularx}
\caption{Calculated metrics using SnykCode.}
\label{table:snykcode_stats}
\end{table}

\begin{table*}[]
\centering
\begin{tabularx}{\textwidth}{XXXXXX}
\multicolumn{6}{c}{\textbf{GPT-4.1}} \\
\hline
\multirow{2}{*}{\textbf{Project}} & 
\multirow{2}{*}{\textbf{False Positive}} & 
\multirow{2}{*}{\textbf{False Negative}} & 
\multirow{2}{*}{\textbf{Precision}} & 
\multirow{2}{*}{\textbf{Recall}} & 
\multirow{2}{*}{\textbf{F1 Score}} \\
& & & & & \\
\hline
S01 & 5 & 0 & 0.615 & 1.000 & 0.762 \\
S02 & 0 & 0 & 1.000 & 1.000 & 1.000 \\
S03 & 2 & 0 & 0.750 & 1.000 & 0.857 \\
S04 & 1 & 0 & 0.500 & 1.000 & 0.667 \\
S05 & 0 & 2 & 1.000 & 0.667 & 0.800 \\
S06 & 3 & 1 & 0.700 & 0.875 & 0.778 \\
S07 & 4 & 2 & 0.733 & 0.846 & 0.786 \\
S08 & 2 & 2 & 0.714 & 0.714 & 0.714 \\
S09 & 1 & 0 & 0.889 & 1.000 & 0.941 \\
S10 & 1 & 1 & 0.667 & 0.667 & 0.667 \\
\hline
\end{tabularx}
\caption{Calculated metrics for each project using GPT-4.1}
\label{table:gpt4_1_stats}
\end{table*}

\begin{table*}[]
\centering
\begin{tabularx}{\textwidth}{XXXXXX}
\multicolumn{6}{c}{\textbf{Mistral Large}} \\
\hline
\multirow{2}{*}{\textbf{Project}} & 
\multirow{2}{*}{\textbf{False Positive}} & 
\multirow{2}{*}{\textbf{False Negative}} & 
\multirow{2}{*}{\textbf{Precision}} & 
\multirow{2}{*}{\textbf{Recall}} & 
\multirow{2}{*}{\textbf{F1 Score}} \\
& & & & & \\
\hline
S01 & 4 & 1 & 0.636 & 0.875 & 0.737 \\
S02 & 1 & 1 & 0.667 & 0.667 & 0.667 \\
S03 & 2 & 0 & 0.750 & 1.000 & 0.857 \\
S04 & 2 & 0 & 0.333 & 1.000 & 0.500 \\
S05 & 1 & 2 & 0.800 & 0.667 & 0.727 \\
S06 & 2 & 2 & 0.750 & 0.750 & 0.750 \\
S07 & 1 & 5 & 0.889 & 0.615 & 0.727 \\
S08 & 0 & 2 & 1.000 & 0.714 & 0.834 \\
S09 & 0 & 1 & 1.000 & 0.875 & 0.934 \\
S10 & 0 & 1 & 1.000 & 0.667 & 0.800 \\
\hline
\end{tabularx}
\caption{Calculated metrics for each project using Mistral Large}
\label{table:mistral_large_stats}
\end{table*}

\begin{table*}[]
\centering
\begin{tabularx}{\textwidth}{XXXXXX}
\multicolumn{6}{c}{\textbf{DeepSeek V3}} \\
\hline
\multirow{2}{*}{\textbf{Project}} & 
\multirow{2}{*}{\textbf{False Positive}} & 
\multirow{2}{*}{\textbf{False Negative}} & 
\multirow{2}{*}{\textbf{Precision}} & 
\multirow{2}{*}{\textbf{Recall}} & 
\multirow{2}{*}{\textbf{F1 Score}} \\
& & & & & \\
\hline
S01 & 4 & 1 & 0.636 & 0.875 & 0.737 \\
S02 & 0 & 0 & 1.000 & 1.000 & 1.000 \\
S03 & 2 & 0 & 0.750 & 1.000 & 0.857 \\
S04 & 4 & 0 & 0.200 & 1.000 & 0.334 \\
S05 & 1 & 1 & 0.833 & 0.833 & 0.834 \\
S06 & 2 & 2 & 0.750 & 0.750 & 0.750 \\
S07 & 2 & 5 & 0.800 & 0.615 & 0.696 \\
S08 & 1 & 1 & 0.857 & 0.857 & 0.857 \\
S09 & 0 & 1 & 1.000 & 0.875 & 0.934 \\
S10 & 3 & 1 & 0.400 & 0.667 & 0.500 \\
\hline
\end{tabularx}
\caption{Calculated metrics for each project using DeepSeek~V3}
\label{table:deepseek_v3_stats}
\end{table*}

\begin{table}[h!] 
\centering 
\begin{tabularx}{\linewidth}{XXXX} \hline \textbf{Tool} & \textbf{Avg. Precision} & \textbf{Avg. Recall} & \textbf{Avg. F1 Score} \\ \hline 
SonarQube & 0.567 & 0.187 & 0.260 \\ 
CodeQL & 0.625 & 0.284 & 0.386 \\ 
SnykCode & 0.686 & 0.523 & 0.546 \\ 
GPT-4.1 & 0.757 & 0.877 & 0.797 \\ 
Mistral Large & 0.783 & 0.783 & 0.753 \\ 
DeepSeek V3 & 0.723 & 0.847 & 0.750 \\ \hline 
\end{tabularx} 
\caption{Average Precision, Recall and F1 Score for each tool and model} \label{table:avg_metrics_summary} 
\end{table}

\subsection{Comparison}

\begin{figure}[H]
    \centering
    \includegraphics[width=1\linewidth]{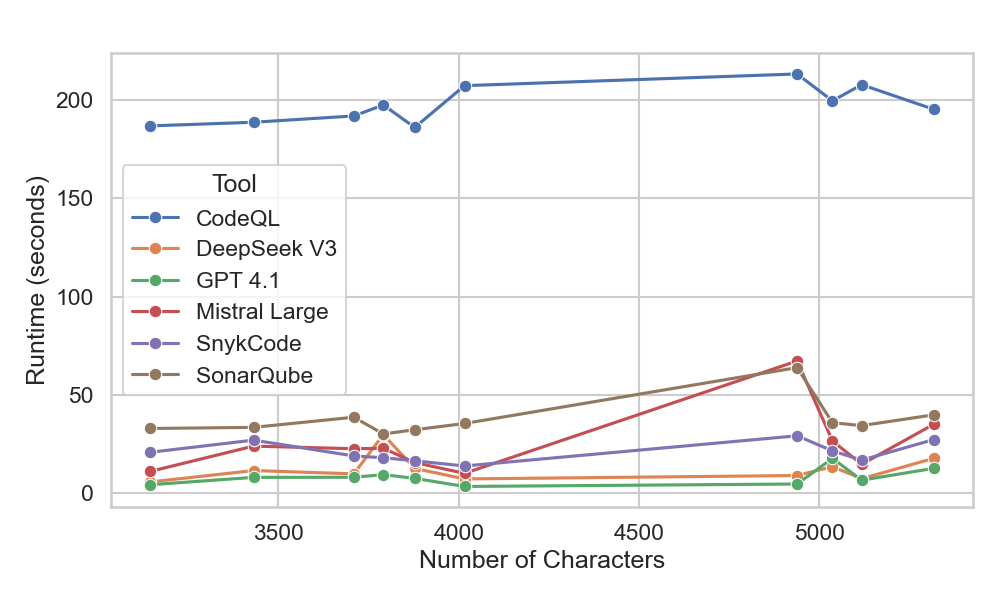}
    \caption{Tool Runtime vs. Number of Characters}
    \label{fig:runtime-vs-characters}
\end{figure}

Figure~\ref{fig:runtime-vs-characters} illustrates the analysis runtime as a function of project size, measured by the number of characters. As shown, for a typical change size as reported by \cite{ferreira2022commits}, the number of characters does not substantially impact the analysis time. However, a clear outlier is \textit{CodeQL}, consistently demonstrating significantly longer runtimes within this range. This behaviour can probably be attributed to its architecture and operational design. Unlike lighter tools tailored for rapid feedback, CodeQL is primarily intended for continuous integration and deployment (CI/CD) environments, where more extensive computational resources are allocated, and longer processing times are acceptable in exchange for deeper and more thorough analysis.

\newpage

\subsection{Number of False Positives}

\begin{figure}[H]
    \centering
    \includegraphics[width=1\linewidth]{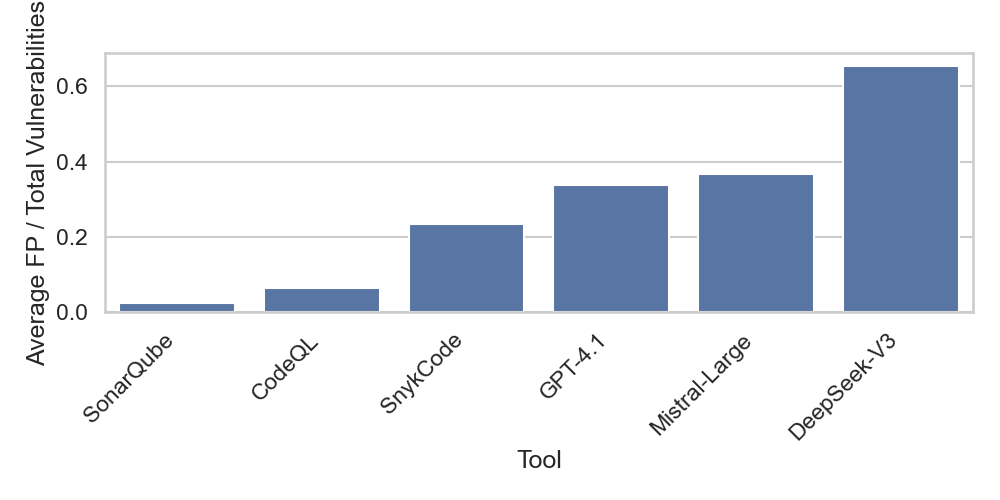}
    \caption{Average FP to Total Found Vulnerabilities Ratio by Tool}
    \label{fig:fp-ratio-comparison}
\end{figure}

Figure~\ref{fig:fp-ratio-comparison} illustrates the juxtaposition of the average false positive rate (FP) with the total number of vulnerabilities found by each tool and the large language model. A lower ratio indicates higher precision in vulnerability detection, reflecting a lower percentage of incorrect alerts than the total number of reported vulnerabilities. 

As shown in the figure, \textit{SonarQube} and \textit{CodeQL} have the lowest FP ratios, suggesting that these traditional tools are more accurate and targeted in their analysis, generating fewer false alerts and highlighting vulnerabilities with greater reliability. Their performance indicates a level of reliability that can be especially appreciated in software development environments where accuracy and confidence in results are essential.

On the other hand, \textit{DeepSeek V3} shows the highest FP ratio in the investigated solutions. This behaviour may result in the need to increase the effort in verification on the part of developers, who must spend more time manually reviewing and validating the analysis results. This suggests that while the tool may be accurate in its assessments, it may also overwhelm users with alerts that may not correspond to actual threats.

It is also important to note that the performance of \textit{SnykCode} is closely aligned with that of LLM-based tools, with its FP ratio between those of traditional analysers and large language models. This similarity may be due to its hybrid analytical approach, which combines machine learning algorithms with techniques commonly employed in static analysis \cite{snykCodeDocs}. The fusion of these methods probably shapes its detection behaviour, resulting in a reporting pattern that reflects characteristics of both traditional and AI-enhanced approaches.

\subsection{Average F1 Score}

\begin{figure}[H]
    \centering
    \includegraphics[width=1\linewidth]{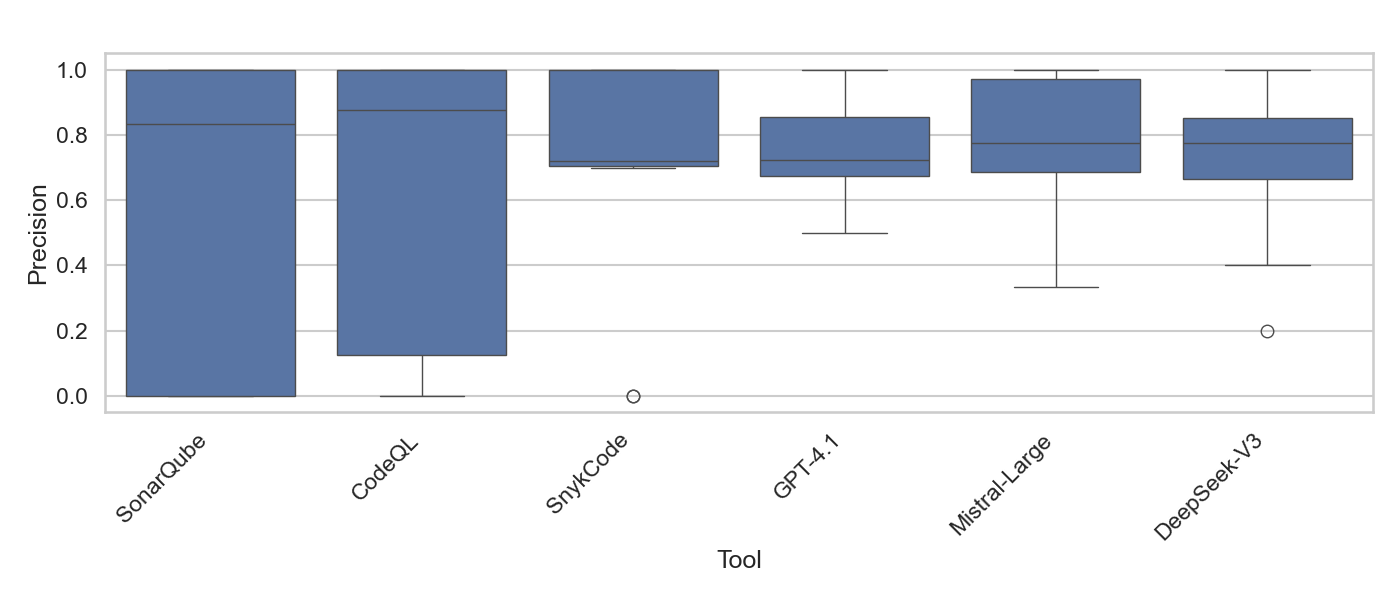}
    \caption{Comparison of effectiveness (Precision)}
    \label{fig:precision-comparison}
\end{figure}

\begin{figure}[H]
    \centering
    \includegraphics[width=1\linewidth]{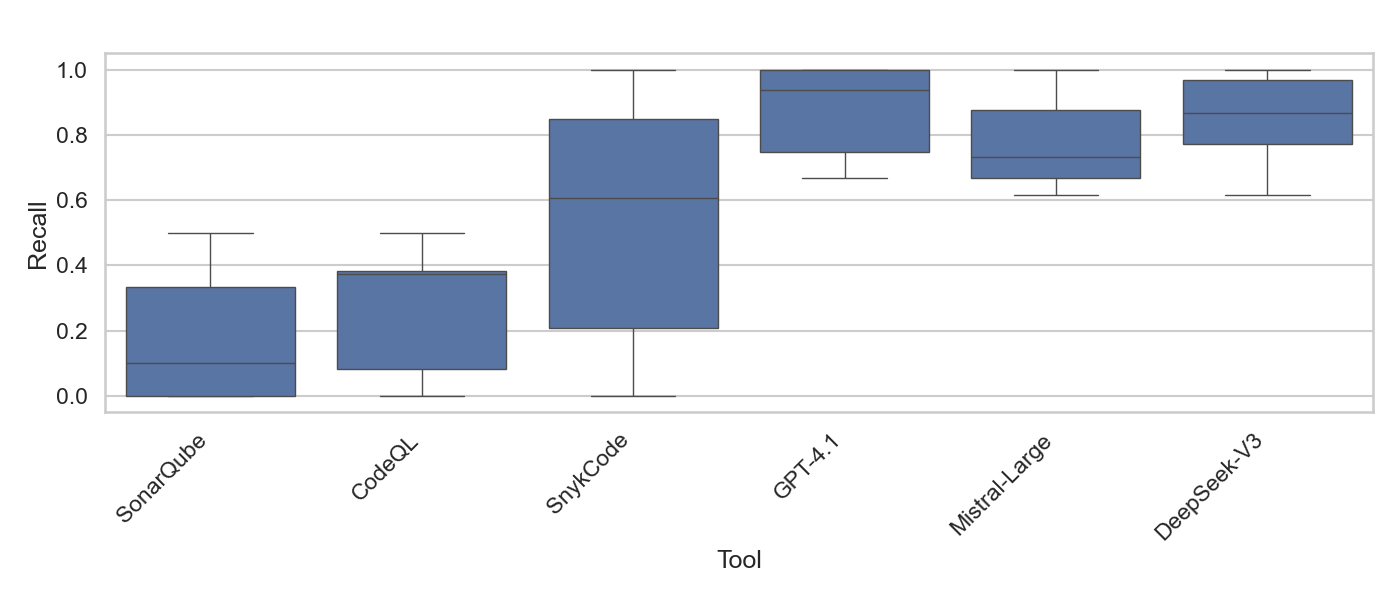}
    \caption{Comparison of Effectiveness (Recall)}
    \label{fig:recall-comparison}
\end{figure}

\begin{figure}[H]
    \centering
    \includegraphics[width=1\linewidth]{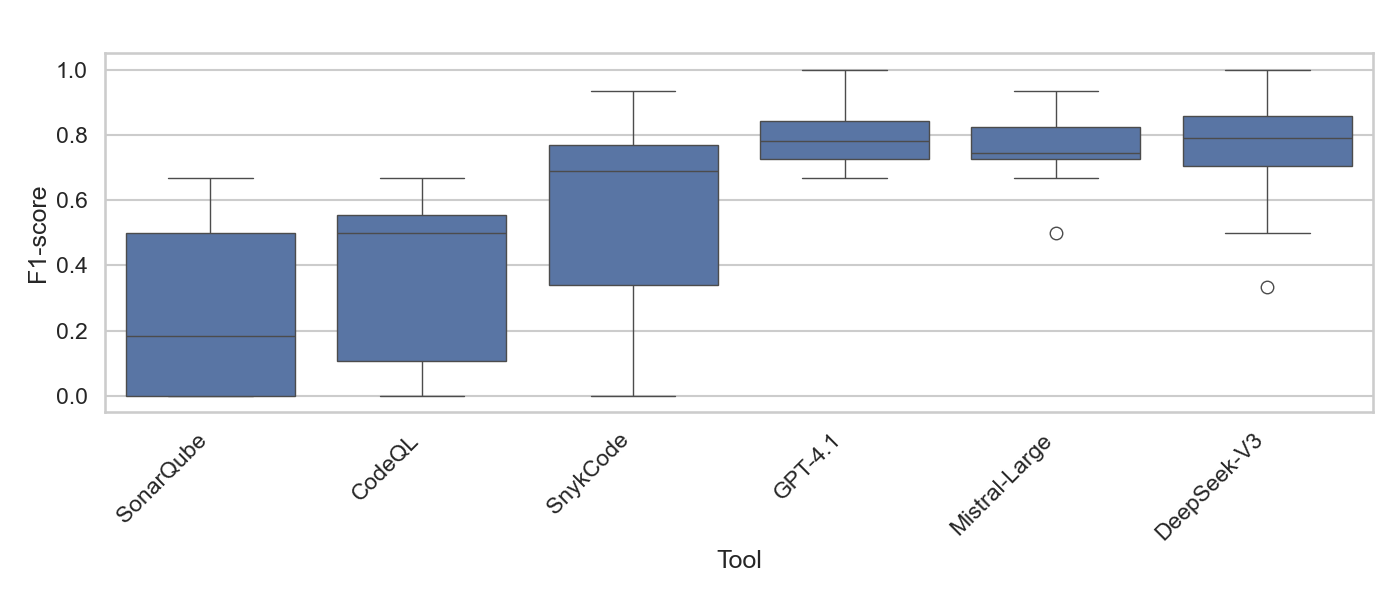}
    \caption{Comparison of Effectiveness (F1 score)}
    \label{fig:f1-score-comparison}
\end{figure}

Figure~\ref{fig:f1-score-comparison} builds on the insights from Figure~\ref{fig:precision-comparison} Figure~\ref{fig:recall-comparison} and compares the effectiveness of each tool using the F1 score. The FP ratio highlights the effectiveness of a tool in isolation. Still, the F1 score provides a more comprehensive measure that balances both the precision of correct detections and the completeness of the tool in recognising issues.

Figure~\ref{fig:precision-comparison} and Figure~\ref{fig:recall-comparison} show that \textit{SnykCode} performs better than other traditional static analysis tools in the F1 score and competes with systems based on large language models, further validating the effectiveness of its hybrid analysis strategy. The tool created by \textit{Snyk Company} achieved the highest average F1 score of 0.55; however, it should be noted that it has two outliers for projects \textit{S04} and \textit{S10}, which can be easily seen earlier, in Figure~\ref{fig:precision-comparison}. In contrast, although strong in precision as shown earlier, \textit{SonarQube} and \textit{CodeQL} have more limited recall, leading to greater variance and lower medians in their F1 score distributions.

\textit{GPT-4.1}, \textit{Mistral-Large} and \textit{DeepSeek V3} achieve higher F1 scores than traditional tools such as \textit{SonarQube} and \textit{CodeQL}. Despite its high false-positive rate, \textit{DeepSeek V3} performs well in terms of F1 score, suggesting that this compensates for its lower precision shown in Figure~\ref{fig:runtime-vs-characters}. It is worth mentioning that with the use of large language models, outliers can also be observed for \textit{Mistral-Large} and \textit{DeepSeek V3} in the \textit{S04} project. These can be observed in Figure~\ref{fig:f1-score-comparison}.

\section{Discussion, Limitations, and Future Work}\label{sec:discussion}

One of the fields found in the SARIF schema is \textit{region}, which reports the location of vulnerabilities found with line and column precision in a given file \cite{sarif-schema}. All of the large language models tested were unable to determine it correctly. All results correctly identified the file, but the incorrect location was always reported. Upon further analysis, no correlation was found in these errors. They are likely caused by one of the steps the text undergoes before processing: BPE tokenisation, which transformer-based models use \cite{chai2024tokenization}. This is a known and common problem. It can be a major limitation in building systems that require precisely determining error locations.

CodeQL achieved the worst results of all the static code analysis tools tested, regarding the time required for analysis. This may be due to the intended use of this tool, which was designed for professional applications that use many predefined rules that the user can define. The need to load them into memory and process them can be a potential reason for these results.

A well-known problem that occurs when using large language models is hallucinations. Is it a situation where the generated response is deviated, based neither on training data nor input from the user \cite{bang2025hallulens,szandala2025aiops}. Such an answer only appears to be correct, but it contains false information. Table~\ref{table:hallucinations} contains examples of such findings. In each case, the latest available versions of the libraries are used in the analysed projects, but the models consider them obsolete. This situation cannot occur with classical code analysis tools, as they usually have a list of deprecated dependencies updated regularly.

\begin{table*}[]
\centering
\begin{tabularx}{\linewidth}{ssXg}
\textbf{Project} & \textbf{Model} & \textbf{Rule ID} & \textbf{Message} \\
\hline
S02 & Mistral Large & VULNERABLE\_LIBRARY & The version '9.0.5' of the 'Microsoft.Data.Sqlite' package is outdated. Consider updating to a more recent version. \\
\hline
S03 & Mistral Large & WEAK\_ALGORITHM & Using an outdated version of Microsoft.Data.SQLite library. \\
\hline
S06 & GPT 4.1 & OUTDATED\_DEPENDENCY & The Newtonsoft.Json package is used (version 13.0.3). Ensure to update it against security vulnerabilities. \\
\hline
\end{tabularx}
\caption{Hallucinated findings by project and model}
\label{table:hallucinations}
\end{table*}

When analysing the results, false positives and false negatives were considered. From the end user's point of view, false negatives are more dangerous, as they translate into overlooking real vulnerabilities that could affect the system's functioning. However, it is worth noting that false positive results also have negative effects. The primary one is that developers pay less attention to the results of the analysis. As a consequence, this can lead to the ignored real issues \cite{cheirdari2018analyzing}.

\subsection{Future Works}

Although this thesis provides a comparative analysis of traditional static analysis tools such as SonarQube, CodeQL and SnykCode and the large language models GPT-4.1, Mistral Large and DeepSeek~V3 in the context of vulnerability detection, several directions remain open for further exploration and development. One aspect worth exploring is prompt engineering, i.e. manipulating the content of queries that are sent to the models and their impact on the results received. 

Another direction could be an attempt to create a hybrid solution that uses both classical tools and language models. These approaches can be used to prioritise and verify vulnerabilities found. This could contribute to reducing the weaknesses of both approaches.

\FloatBarrier
\section{Conclusion}\label{sec:conclusion} 
After analysing all projects using the selected static code analysis tools and large language models, conclusions can be drawn. Large language models perform well in detecting code vulnerabilities, as seen from comparing the F1 score values shown in Figure~\ref{fig:f1-score-comparison}. The ability to analyse the entire context of the code, as opposed to traditional tools that rely on defined patterns, can be considered the reason. It is also worth noting that the increased probability of reporting false positives comes with the greater ability to detect vulnerabilities. This is shown in Figure~\ref{fig:fp-ratio-comparison}. Another important factor when considering large language models as code analysis tools is their inability to provide the exact location of the vulnerabilities found. When developing more elaborate code analysis systems, the failure to provide location information when the SARIF format is used may preclude using large language models. 

In conclusion, classical tools for static code analysis are recommended in situations requiring reliability and precision, such as software audits, which are necessary for systems that must be reliable and used in critical sectors. In such situations, the definition of specific patterns is often required. Solutions based on large language models have a higher sensitivity, but generate more false-positive results. They can be used during the development process to ensure that developers working on the system are aware of defects.




\bibliographystyle{unsrt}
\bibliography{bibliography}




\begin{IEEEbiography}
[{\includegraphics[width=1in,height=1.25in,clip,keepaspectratio]{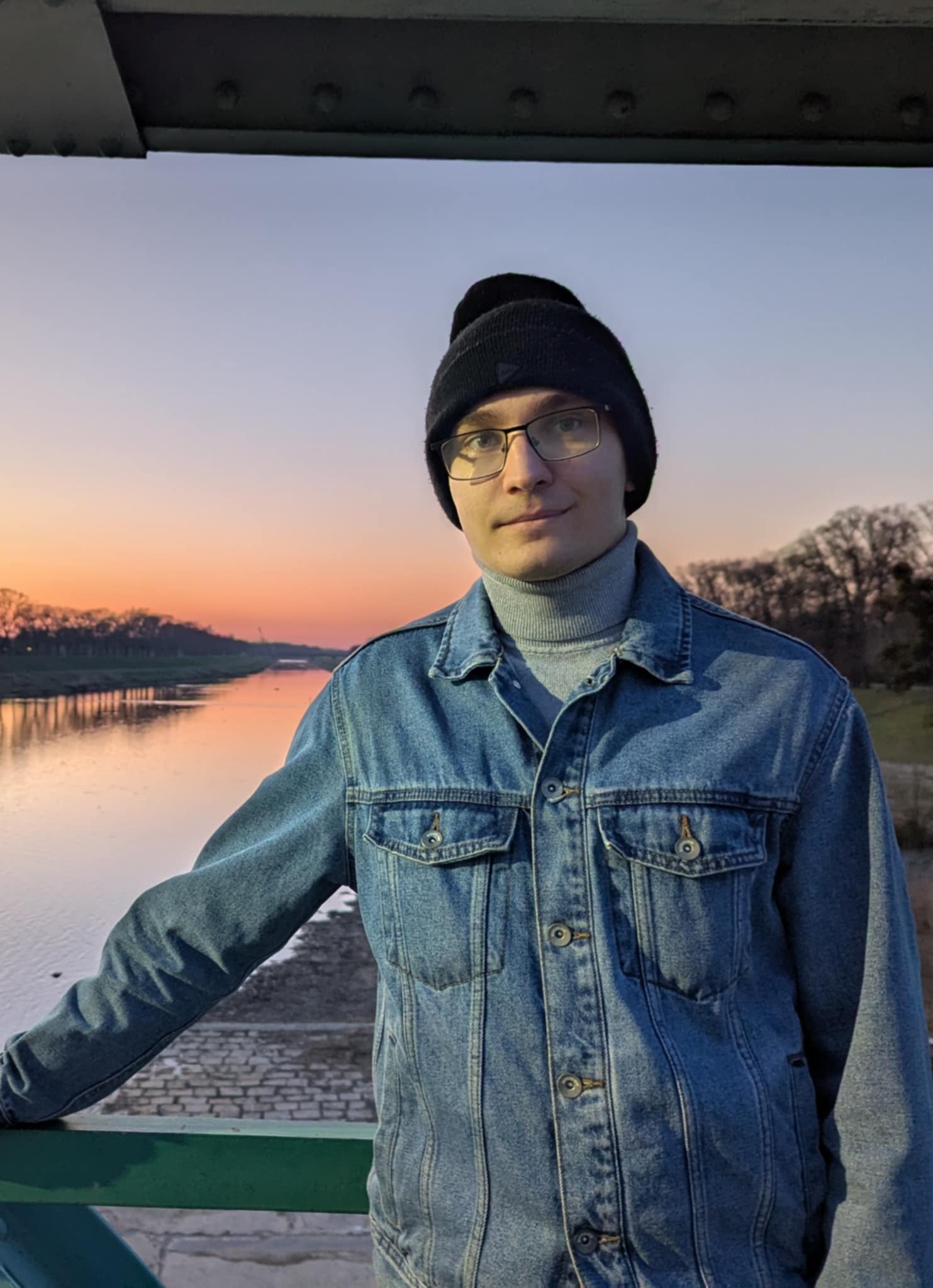}}]{Damian Gnieciak} a WIT graduate with a master’s degree in Computer Science and a software engineer with a professional focus on cloud technologies, is currently developing a specialization in Microsoft Azure. He is particularly interested in the design and implementation of scalable, secure, and highly available solutions.
\end{IEEEbiography}

\begin{IEEEbiography}
[{\includegraphics[width=1in,height=1.25in,clip,keepaspectratio]{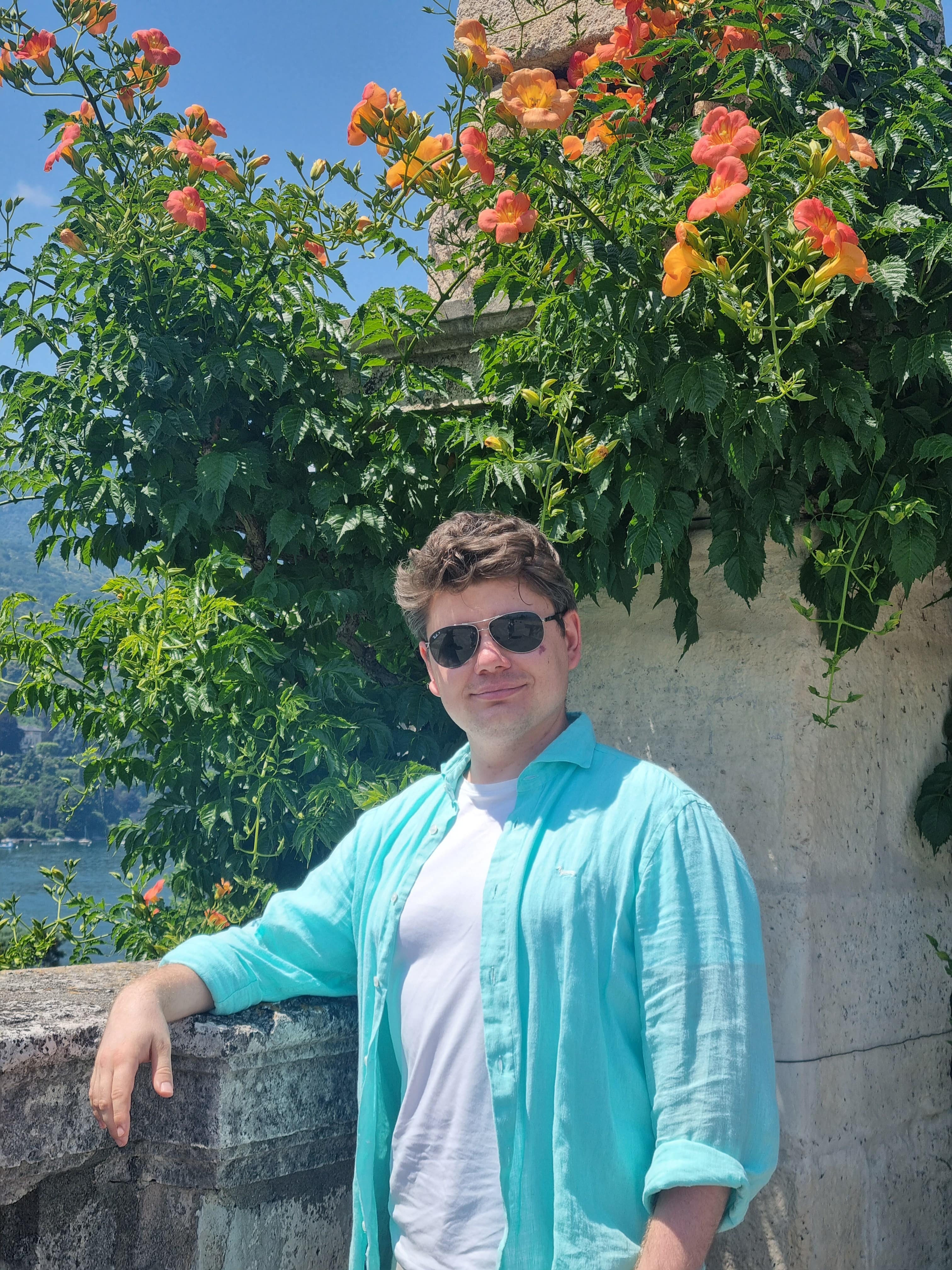}}]{Tomasz Szandala} a Ph.D. graduate in Computer Science (2022), is a PostDoc researcher at the Scuola Universitaria Professionale della Svizzera italiana in Lugano, supported by an ESKAS scholarship. His academic work centers on applied computer science and DevOps. In addition to academic pursuits, Tomasz is a certified DevOps engineer with expertise in Kubernetes and Google Cloud Platform, frequently publishing industrial papers that bridge theoretical and practical insights across his academic and professional experience.
\end{IEEEbiography}

\EOD

\end{document}